\newif\ifrevtex
\IfFileExists{revtex4-2.cls}{\revtextrue}{\revtexfalse}
\ifrevtex
  \documentclass[aps,prd,twocolumn,superscriptaddress,nofootinbib,floatfix]{revtex4-2}
\else
  \documentclass[11pt,a4paper]{article}
  \usepackage[margin=2.4cm]{geometry}
  \usepackage[font=small,labelfont=bf]{caption}
\fi
\usepackage[utf8]{inputenc}
\ifrevtex\usepackage[T1]{fontenc}\fi
\usepackage{amsmath,amssymb}
\usepackage{graphicx}
\usepackage{booktabs}
\usepackage{etoolbox}
\usepackage[colorlinks=true,linkcolor=blue,citecolor=blue,urlcolor=blue]{hyperref}

\newcommand{\Nch}{\langle dN_{\rm ch}/d\eta\rangle}
\newcommand{\HL}{{}^{3}_{\Lambda}\mathrm{H}}
\newcommand{\He}{{}^{3}\mathrm{He}}
\ifrevtex\else\AtBeginEnvironment{table}{\centering}\fi
\ifrevtex\newcommand{\colfig}{0.95\columnwidth}\newcommand{\widefig}{0.95\textwidth}\else\newcommand{\colfig}{0.55\textwidth}\newcommand{\widefig}{0.95\textwidth}\fi

\newcommand{\papertitle}{Exact flavor conservation at the quark level in causal clusters:\\
strange particles, protons and light nuclei in small collision systems}
\newcommand{\paperabstract}{The suppression of strange hadrons in pp and p--Pb collisions at
the LHC, including the $\phi$ meson, has recently been described by the exact conservation
of strangeness imposed on the quarks of a causally connected cluster before
hadronization. Here we extend this approach to the light flavors. The numbers of up
and down quarks and antiquarks are conserved in the same cluster, and the suppression of
a baryon or a nucleus depends only on how many light quarks of each flavor it contains.
With the strangeness sector fixed by the previous analysis, a single constant, the ratio
$r$ of light to strange quark abundances, determines the multiplicity dependence of
protons, deuterons, $\He$ and hypertriton in pp and p--Pb collisions. Including the finite
size of the nuclei, one obtains a very good fit of the data for $r=3.76\pm0.38$, i.e.\
strange quarks about 0.27 times as abundant as light quarks per flavor, close to the value
known from string fragmentation.}

\begin{document}

\ifrevtex
\title{\papertitle}
\author{P.~Castorina}
\affiliation{INFN, Sezione di Catania, Catania, Italy}
\affiliation{Faculty of Mathematics and Physics, Charles University, Prague, Czech Republic}
\author{H.~Satz}
\affiliation{Fakult\"at f\"ur Physik, Universit\"at Bielefeld, Bielefeld, Germany}
\date{\today}
\begin{abstract}\paperabstract\end{abstract}
\maketitle
\else
\title{\papertitle}
\author{P.~Castorina$^{1,2}$ and H.~Satz$^{3}$\\[2pt]
\small $^{1}$INFN, Sezione di Catania, Catania, Italy\\
\small $^{2}$Faculty of Mathematics and Physics, Charles University, Prague, Czech Republic\\
\small $^{3}$Fakult\"at f\"ur Physik, Universit\"at Bielefeld, Bielefeld, Germany}
\date{\today}
\maketitle
\begin{abstract}\noindent\paperabstract\end{abstract}
\fi

\section{Introduction}

In central nucleus-nucleus collisions the abundances of hadrons and light nuclei are
those of an ideal gas of resonances at a temperature $T\simeq155$
MeV~\cite{Andronic:2017pug}. In pp and p--Pb collisions the ALICE Collaboration has shown
that the yields of strange hadrons, relative to pions, rise smoothly with the
charged-particle multiplicity towards the values of central Pb--Pb collisions, more
strongly for hadrons with more strange quarks~\cite{ALICE:2016fzo,ALICE:2015mpp,
ALICE:2019avo}. Light nuclei show a similar, even stronger, rise: the deuteron-to-proton
ratio increases by a factor four from the lowest to the highest multiplicities in pp
collisions~\cite{ALICE:2021mfm,ALICE:2021ovi}, while the proton-to-pion ratio changes
by little more than 10\%~\cite{ALICE:2018pal}.

The suppression of strange hadrons is usually described by the exact conservation of
strangeness in a finite volume, imposed on hadrons~\cite{Rafelski:1980gk,
Hagedorn:1984uy,Hamieh:2000tk,Kraus:2007hf,Kraus:2008fh}. In a previous
letter~\cite{CS:PLB} we showed that the conservation law should instead be imposed on
the strange quarks of a causally connected cluster, before hadronization. For hadrons
with open strangeness the two procedures give the same hierarchy of suppression
factors. For the $\phi$ meson, which has no net strangeness, at the hadron
level there is no suppression, whereas at
the quark level one can distinguish the combinations of a strange quark with the
antiquark produced together with it from those with the antiquark of another pair, and
the different quantum numbers of the two lead to a suppression of the $\phi$ in agreement
with the data.

The same distinction connects strangeness production to the spin correlation of
$\Lambda\bar\Lambda$ pairs recently observed by the STAR
Collaboration~\cite{STAR:2025njp}, interpreted as the survival of the parallel spins of
an $s\bar s$ pair created from the vacuum. A $\Lambda$ and a $\bar\Lambda$ carry this
correlation only if their strange quark and antiquark come from the same pair. The
fraction of such combinations is fixed by the same pair distribution that determines the
suppression of the strange hadrons; it decreases with multiplicity, and the spin
correlation is predicted to decrease in the same proportion~\cite{CS:PLB}. Strangeness
suppression, the $\phi$ meson and the hyperon spin correlations are thus three
consequences of a single mechanism.

In this paper we extend the same description to the light flavors. Up and down quarks
are conserved in the same cluster, and the suppression of protons and light nuclei
follows from the numbers of light quarks they contain. Since the strangeness sector is
fixed by the previous analysis, only one new constant appears: the ratio of light to
strange quark abundances. For nuclei we add the overlap between the bound state and the
cluster, which is relevant because their size is comparable with that of the cluster.

In Sec.~\ref{sec:strange} we summarize the results for strangeness, including the
spin correlation of hyperon pairs. In Sec.~\ref{sec:light} we extend the formalism to
light quarks, including the finite size of the nuclei.
Section~\ref{sec:fit} presents the comparison with the data on protons and light nuclei,
Sec.~\ref{sec:cluster} discusses the cluster size and the meaning of the new constant,
and Sec.~\ref{sec:disc} contains the discussion and the conclusions.

\section{Strangeness at the quark level}\label{sec:strange}

We first recall the framework and the results of Ref.~\cite{CS:PLB}, to which we refer
for the derivations.

\subsection{Causal clusters and pair distribution}

Consider a boost-invariant expansion in which partons materialize at the proper time
$\tau_q$ and hadrons appear at $\tau_h$. Two points can exchange a signal during this
interval only if their proper longitudinal distance does not exceed~\cite{Castorina:2013mba}
\begin{equation}
d=\sqrt{\frac{\tau_h}{\tau_q}}\,(\tau_h-\tau_q). \label{eq:d}
\end{equation}
Since flavor is conserved locally, quarks appear together with their antiquarks, whether
they come from string breaking or from gluon splitting, and the two members of a pair
remain within the distance $d$ until hadronization. The causally connected region, which
we approximate by a sphere of radius $R=d/2$ and volume $V$, therefore contains as many
quarks as antiquarks of every flavor. We refer to it as a cluster.

We assume that the production in a cluster is statistical from the outset: every
configuration compatible with the conservation law is populated according to its
statistical weight. If each quark and each antiquark of flavor $f$ has the weight
$x_f/2$, the probability of finding $n$ pairs is~\cite{CS:PLB}
\begin{equation}
P(n)=\frac{1}{I_0(x_f)}\,\frac{(x_f/2)^{2n}}{(n!)^2}, \label{eq:P}
\end{equation}
where $I_k$ denotes the modified Bessel function of order $k$. The parameter $x_f$ is
the average number of quarks plus antiquarks of flavor $f$ that a cluster of the same
volume would contain if the constraint were absent; for a Boltzmann gas at temperature
$T$,
\begin{equation}
x_f=2Vn_f,\qquad n_f=\frac{g}{2\pi^2}\,m_f^2\,T\,K_2\!\Big(\frac{m_f}{T}\Big),
\label{eq:x}
\end{equation}
with $g=6$ spin and color states. The quantity that enters the hadron yields is the
factorial moment
\begin{equation}
\big\langle n(n-1)\cdots(n-k+1)\big\rangle=\Big(\frac{x_f}{2}\Big)^{k}
\frac{I_k(x_f)}{I_0(x_f)} , \label{eq:fact}
\end{equation}
i.e.\ the number of ways of choosing $k$ distinct quarks of flavor $f$ in the cluster.
Without the constraint this number would be $(x_f/2)^k$. A hadron that needs $k$ quarks
(or $k$ antiquarks) of flavor $f$ is therefore reduced, with respect to a very large
cluster, by the factor $I_k(x_f)/I_0(x_f)$.

\subsection{Open and hidden strangeness}

For strange quarks we write $S_k\equiv I_k(x_s)/I_0(x_s)$ and then Kaons and $\Lambda$  are
reduced by $S_1$, the $\Xi$ by $S_2$ and the $\Omega$ by $S_3$. The $\phi$ meson requires
one strange quark and one strange antiquark, and the relevant number is
$\langle n^2\rangle$, which is exactly $(x_s/2)^2$: the conservation law by itself leaves
hidden strangeness unsuppressed. However, the combinations of a quark with an antiquark are of two kinds,
\begin{equation}
\langle n^2\rangle=\langle n(n-1)\rangle+\langle n\rangle
=\Big(\frac{x_s}{2}\Big)^2\big[S_2+(1-S_2)\big], \label{eq:split}
\end{equation}
the first term counting quarks and antiquarks of different pairs, the second a quark
with its own partner. The two kinds differ in their quantum numbers. A pair produced
from the vacuum is in the state $J^{PC}=0^{++}$ and cannot become a $\phi$, whose quantum
numbers are $1^{--}$; a pair produced by a gluon has the right $J^{PC}$ but carries
color, and must neutralize it before it can form a $\phi$. Denoting by $w$ the fraction
of same-pair combinations that end up in a $\phi$, the suppression factor of the $\phi$
becomes
\begin{equation}
C_\phi=S_2+w\,(1-S_2). \label{eq:Cphi}
\end{equation}

\subsection{Fit to the strange hadrons}

\begin{figure}[t]
\centering
\includegraphics[width=\colfig]{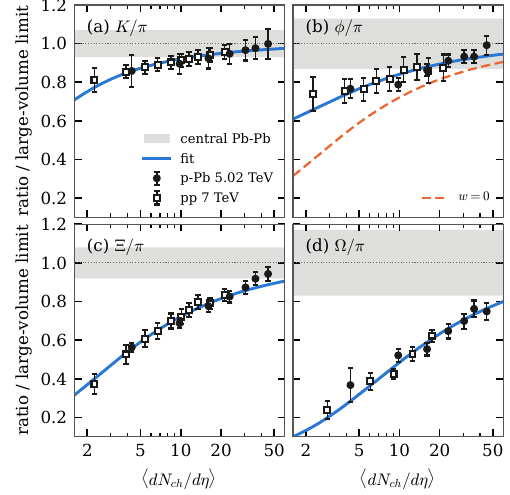}
\caption{Strange hadron to pion ratios in p--Pb collisions at $\sqrt{s_{NN}}=5.02$
TeV~\cite{ALICE:2016sak,ALICE:2015mpp} and in pp collisions at $\sqrt s=7$
TeV~\cite{ALICE:2018pal}, divided by the large-cluster limits and by the fitted
normalization factors, versus charged-particle multiplicity, from Ref.~\cite{CS:PLB}.
Solid lines: $S_1$, $C_\phi$, $S_2$ and $S_3$ with Eq.~(\ref{eq:fitS}). Dashed line in
(b): $w=0$. Shaded bands: central Pb--Pb values with their uncertainties.}
\label{fig:strange}
\end{figure}

In Ref.~\cite{CS:PLB} these factors were compared with ALICE data  on
$K/\pi$, $\phi/\pi$, $\Xi/\pi$ and $\Omega/\pi$, in pp collisions at $\sqrt s=7$
TeV~\cite{ALICE:2018pal} and in p--Pb collisions at $\sqrt{s_{NN}}=5.02$
TeV~\cite{ALICE:2016sak,ALICE:2015mpp}. The ratios expected for a very large cluster were
taken from central Pb--Pb collisions at $\sqrt{s_{NN}}=2.76$ TeV, and each measurement was
given a normalization factor accounting for the part of its systematic uncertainty that
is common to all multiplicity classes. The dependence of the strange-quark content on
the charged-particle multiplicity was parametrized as
\begin{equation}
x_s=a\,N^{b},\qquad N\equiv\Nch , \label{eq:xs}
\end{equation}
and the fit gave
\begin{equation}
a=1.54\pm0.13,\qquad b=0.63\pm0.06,\qquad w=0.43\pm0.09, \label{eq:fitS}
\end{equation}
with $\chi^2=37.9$ for 59 degrees of freedom. The pp and p--Pb data lie on the same
curves (Fig.~\ref{fig:strange}). The two extreme assumptions for the $\phi$, no
suppression ($w=1$) and suppression as a doubly strange hadron ($w=0$), are excluded by
5.7 and 4.8 standard deviations. The average number of strange quarks and antiquarks per
cluster grows from 2.6 at $N=2.3$ to 17 at $N=45$. As an independent check, the $\phi$
yield in p--Pb collisions at $\sqrt{s_{NN}}=8.16$~TeV~\cite{ALICE:2021uyz}, which was not
included in the fit, follows the predicted multiplicity dependence. Throughout this paper
$a$ and $b$ are kept at the values of Eq.~(\ref{eq:fitS}).

\subsection{Spin correlation of \texorpdfstring{$\Lambda\bar\Lambda$}{Lambda anti-Lambda} pairs}

The STAR Collaboration has observed a relative polarization of $0.181\pm0.035\pm0.022$ for
$\Lambda\bar\Lambda$ pairs that are close in rapidity and azimuthal angle, in pp
collisions at $\sqrt s=200$ GeV, while no correlation is found for distant pairs or for
$\Lambda\Lambda$ pairs~\cite{STAR:2025njp}. The effect is attributed to the parallel spins
of an $s\bar s$ pair produced from the vacuum, which survive hadronization.

\begin{figure}[t]
\centering
\includegraphics[width=\colfig]{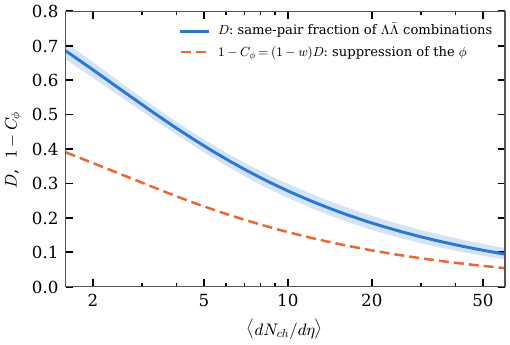}
\caption{Fraction $D$ of $\Lambda\bar\Lambda$ combinations whose strange quark and
antiquark were produced in the same pair, Eq.~(\ref{eq:D}), with its $1\sigma$ band
(solid line), and suppression of the $\phi$, $1-C_\phi$, Eq.~(\ref{eq:rel}) (dashed
line), as functions of multiplicity. From Ref.~\cite{CS:PLB}.}
\label{fig:D}
\end{figure}

Within a cluster, the spin correlation can be carried only by a $\Lambda$ and a
$\bar\Lambda$ whose strange quark and antiquark were produced together; this holds also
for pairs produced by a gluon, whose spins are parallel as well. The fraction of such
combinations is the second term of Eq.~(\ref{eq:split}) divided by the total,
\begin{equation}
D=\frac{\langle n\rangle}{\langle n^2\rangle}=\frac{2}{x_s}\,S_1=1-S_2 , \label{eq:D}
\end{equation}
where we used $I_0(x)-I_2(x)=(2/x)\,I_1(x)$. Comparing with Eq.~(\ref{eq:Cphi}) gives
\begin{equation}
1-C_\phi=(1-w)\,D . \label{eq:rel}
\end{equation}
The combinations that preserve the spin correlation are the same ones that, in a
fraction $1-w$, fail to produce a $\phi$. The dilution of the spin correlation and the
suppression of the $\phi$ are therefore proportional at every multiplicity. With the
parameters of Eq.~(\ref{eq:fitS}), $D$ decreases from $0.44\pm0.02$ at $N=4.3$ to
$0.21\pm0.02$ at $N=16$ and $0.11\pm0.02$ at $N=45$ (Fig.~\ref{fig:D}).

The model thus predicts that the $\Lambda\bar\Lambda$ spin correlation falls with
multiplicity, and disappears in central nucleus-nucleus collisions. The
transfer of polarization from the strange quark to the $\Lambda$, and the dilution due to
feed-down, multiply the correlation by a common factor that largely cancels when two
multiplicities are compared; the dependence on multiplicity is a cleaner test than the
absolute magnitude.

Indeed, as for the magnitude, pp collisions at $\sqrt s=200$ GeV correspond to
$N\simeq3$~\cite{STAR:2008med}, where $D=0.53$ if $x_s$ depends on multiplicity only. A
sample of pairs entirely in the spin-triplet state, with complete transfer of the spin to
the $\Lambda$ and the $\bar\Lambda$, would give a relative polarization of $1/3$; using
the SU(6) wave function and the feed-down fractions of PYTHIA, Ref.~\cite{STAR:2025njp}
obtains a maximum of 0.096 instead. With this maximum the dilution leads to 0.05, about
three standard deviations below the data. The measured value, however, already exceeds
this maximum by two standard deviations, so that even an undiluted correlation ($D=1$)
would not reproduce it. With the factor $1/3$ and no feed-down correction the dilution
gives 0.18, close to the measurement. The comparison of absolute values is thus limited by
the polarization transfer rather than by $D$.

\section{Light quarks}\label{sec:light}

The argument of Sec.~\ref{sec:strange} applies to every flavor. Up and down quarks are
also created in pairs, and their numbers are conserved separately in each cluster. We
take the two light flavors to be equally abundant, with the same parameter $x_l$, and
write
\begin{equation}
L_k\equiv\frac{I_k(x_l)}{I_0(x_l)} .
\end{equation}
A hadron or nucleus that contains $k_u$ up, $k_d$ down and $k_s$ strange quarks (or the
same numbers of antiquarks) is suppressed, relative to the limit of a large cluster, by
the product of the factors of the three flavors,
\begin{equation}
F=\frac{I_{k_u}(x_l)}{I_0(x_l)}\,\frac{I_{k_d}(x_l)}{I_0(x_l)}\,
\frac{I_{k_s}(x_s)}{I_0(x_s)} . \label{eq:F}
\end{equation}
The pion $\pi^+$ has $F=L_1^2$, and the ratios to pions considered below follow
from Table~\ref{tab:content}. The light-quark factors are close to one for hadrons,
since $x_l$ is large, but not for nuclei, which contain many light quarks of the same
flavor (the deuteron needs three up and three down quarks from the same cluster).

\begin{table}[htb]
\caption{Quark content $(k_u,k_d,k_s)$ and suppression factor $F$ of
Eq.~(\ref{eq:F}) for the particles considered; the last column is the factor of the
ratio used in the analysis.}
\label{tab:content}
\begin{tabular}{lccc}
\toprule
 & $(k_u,k_d,k_s)$ & $F$ & ratio \\
\midrule
$\pi^+$ & (1,1,0) & $L_1^2$ & -- \\
$p$ & (2,1,0) & $L_2L_1$ & $p/\pi$: $L_2/L_1$ \\
$d$ & (3,3,0) & $L_3^2$ & $d/p$: $L_3^2/(L_2L_1)$ \\
$\He$ & (5,4,0) & $L_5L_4$ & $\He/p$: $L_5L_4/(L_2L_1)$ \\
$\HL$ & (4,4,1) & $L_4^2S_1$ & $\HL/\pi$: $L_4^2S_1/L_1^2$ \\
${}^4$He & (6,6,0) & $L_6^2$ & ${}^4{\rm He}/p$: $L_6^2/(L_2L_1)$ \\
\bottomrule
\end{tabular}
\end{table}

Exact conservation of the light flavors requires that the cluster carry no net baryon
number. At LHC energies the antiproton-to-proton ratio at midrapidity is compatible with
unity~\cite{ALICE:2010hjm}, so that the valence quarks of the colliding hadrons do not
contribute appreciably to the clusters at midrapidity.

Light and strange
quarks are produced in the same clusters, and their abundances are proportional. We
write
\begin{equation}
x_l=r\,x_s , \label{eq:r}
\end{equation}
where $r$ is the ratio of the mean number of up quarks (or down quarks) to that of
strange quarks that a cluster would contain without the constraint. It is the only new
constant: with $x_s$ fixed by Eqs.~(\ref{eq:xs}) and (\ref{eq:fitS}), $x_l$ is determined
for every multiplicity. In a thermal picture $r=n_u/n_s$ is fixed by the quark masses
through Eq.~(\ref{eq:x}); in string fragmentation $1/r$ is the strangeness suppression
factor $\lambda_s$ of the string-breaking probabilities~\cite{Andersson:1983ia,
Bierlich:2022pfr}. We
determine $r$ from the data and return to its meaning in Sec.~\ref{sec:cluster}.

The factors $L_k$ and $S_k$ refer to a single cluster. An event consists of
several clusters along the rapidity axis, each extending over about
one unit of rapidity. A baryon or a nucleus is formed from the quarks of one cluster, and
its suppression is determined by the number of light quarks available in that cluster,
i.e.\ by $x_l$. The clusters of an event are formed under the same conditions and have
the same quark content; the number of clusters per unit of rapidity cancels in the
ratios $p/\pi$, $d/p$ and $\He/p$, which therefore depend only on the content of a single
cluster. This is the same mechanism that applies to strangeness. It differs from the
canonical description at the hadron level, in which baryon number is conserved in a
correlation volume that has to be chosen separately: the volume required by the deuteron
data suppresses protons in pp collisions much more than observed, while at the quark
level a single volume describes both.

Finally, for light nuclei in small systems, the cluster size is comparable with their
radii. Equation~(\ref{eq:F}) counts the quarks available in the cluster, but not
whether the bound state fits in it. For a nucleus of $A$ nucleons with a Gaussian wave
function of size parameter $r_A$, formed in a Gaussian source of radius $R$, the
probability is reduced by~\cite{Scheibl:1998tk,Bellini:2018epz}
\begin{equation}
G_A(R)=\Big[1+\Big(\frac{r_A}{2R}\Big)^2\Big]^{-3(A-1)/2}. \label{eq:G}
\end{equation}
where $R$ is the radius of the cluster, obtained from $x_l$ through
Eq.~(\ref{eq:x}) with a light quark mass of 300 MeV. We use $r_d=3.2$ fm and
$r_{\He}=2.48$ fm~\cite{Bellini:2018epz}; for ${}^4$He we use $r_A\simeq2.0$ fm, obtained
from its measured charge radius~\cite{Krauth:2021foz} with the same relation between $r_A$
and the root-mean-square radius as for $\He$ in Ref.~\cite{Bellini:2018epz}. The hypertriton
is treated as a deuteron and a $\Lambda$, with factor $G_2(R;r_d)\,G_2(R;r_\Lambda)$ and
$r_\Lambda=8.7$ fm, which corresponds to a root-mean-square separation of about 10 fm
between the $\Lambda$ and the deuteron~\cite{Nemura:1999qp,Hildenbrand:2019sgp} and lies
within the range 6.8--14.1 fm given in Ref.~\cite{Bellini:2018epz}. 

\section{Comparison with data}\label{sec:fit}

\subsection{Data and fit}

The proton-to-pion ratio has been
measured in pp collisions at $\sqrt s=7$ TeV in ten multiplicity classes, with $N$ from
2.3 to 21~\cite{ALICE:2018pal}, and in p--Pb collisions at $\sqrt{s_{NN}}=5.02$ TeV in
seven classes, with $N$ from 4.3 to 45~\cite{ALICE:2016sak}. Proton and deuteron yields
are available in pp collisions at $\sqrt s=5.02$ TeV in nine classes, with $N$ from 2.4
to 18.5~\cite{ALICE:2021mfm}, and at 13 TeV in three high-multiplicity classes, with $N$
from 30 to 36~\cite{ALICE:2021ovi}; from these we use the protons per unit of
charged-particle density and the ratio $d/p$. 

The same two analyses provide the
$\He$-to-proton ratio, in four points with $N$ from 4.9 to 31.5. Finally, the hypertriton
yield has been measured in the 40\% highest-multiplicity p--Pb collisions at 5.02
TeV~\cite{ALICE:2021sdc}, at $N=29.4$, and we use it per unit of charged-particle
density.

The large-cluster limits of $d/p$ and $\He/p$ are the values in
central Pb--Pb collisions at 2.76 TeV~\cite{ALICE:2015wav}, $3.6\times10^{-3}$ and
$9.3\times10^{-6}$. For the hypertriton we use the yield measured in the 10\% most central
Pb--Pb collisions at 2.76 TeV~\cite{ALICE:2015oer}, with the branching ratio of 25\% for the
decay into $\He\,\pi^-$ adopted in Ref.~\cite{ALICE:2021sdc}, per unit of charged-particle
density: $1.07\times10^{-7}$. 

As in Ref.~\cite{CS:PLB}, each data set has a normalization
factor, constrained to unity within the uncertainty of its reference value (16, 25 and
30\%).

 The absolute proton yields are not compared with Pb--Pb collisions, where they are
affected by annihilation in the hadronic phase; their normalizations are free, and only
their dependence on multiplicity is used. The strangeness parameters are fixed to
Eq.~(\ref{eq:fitS}), and the only parameter of the model is $r$.

The fit gives
\begin{equation}
r=3.76\pm0.38, \label{eq:fitr}
\end{equation}
with $\chi^2=27.4$ for 42 degrees of freedom. The details of the procedure are given in
the Supplemental Material.

The size factor of Eq.~(\ref{eq:G}) is required. Without it, the suppression of the
nuclei must come entirely from quark counting: the fit then gives $r\simeq2.1$, which
suppresses the protons too much, and $\chi^2$ increases from 27.4 to 50.6, mostly from
the proton data. The effect of the size factor at fixed $r$ is shown in
Fig.~\ref{fig:nuclei}. The result is
stable with respect to the size attributed to the hypertriton: with $r_\Lambda=6.8$ and
10.3 fm, $\chi^2=26.3$ and 28.8, with the same $r$.

\subsection{Protons}\label{sec:protons}

\begin{figure*}[t]
\centering
\includegraphics[width=0.8\textwidth]{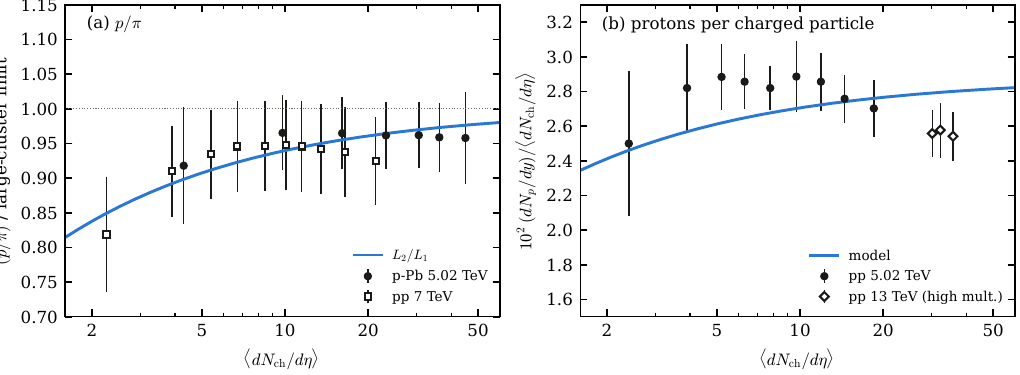}
\caption{(a) Proton-to-pion ratio in p--Pb collisions at $\sqrt{s_{NN}}=5.02$
TeV~\cite{ALICE:2016sak} and in pp collisions at $\sqrt s=7$ TeV~\cite{ALICE:2018pal},
divided by the fitted normalization, versus charged-particle multiplicity. Solid line:
$L_2/L_1$ with Eqs.~(\ref{eq:fitS}) and (\ref{eq:fitr}). (b) Protons per unit of
charged-particle density in pp collisions at 5.02 TeV~\cite{ALICE:2021mfm} and 13
TeV~\cite{ALICE:2021ovi}, with the model.}
\label{fig:protons}
\end{figure*}

The model predicts $p/\pi\propto L_2/L_1$. Since $x_l$ is between 10 and 64 in the range of the
data, the suppression is small: 15\% at the lowest multiplicity in pp collisions and 2\%
at the highest in p--Pb collisions. The proton-to-pion ratio is
constant within a few percent above $N\simeq4$, and decreases by about 12\% in the lowest
multiplicity class of pp collisions at 7 TeV [Fig.~\ref{fig:protons}(a)]. The 17 points
are described with $\chi^2=1.5$.

The number of protons per unit of charged-particle density in pp collisions at 5.02 and
13 TeV is depicted in Fig.~\ref{fig:protons}(b) ($\chi^2=15.6$ for 12 points). In fact,
this ratio, unlike $p/\pi$, includes the dependence on multiplicity of the
number of pions per charged particle, which in p--Pb collisions decreases by about 6\%
over the measured range. The three points at 13 TeV lie
below the extrapolation of those at 5.02 TeV, and give 8.1 of the total.

Most protons come from resonance decays, which are not included in
Table~\ref{tab:content}. Their effect can be estimated roughly. About 45\% of the
protons come from $\Delta$ decays; the $\Delta^{++}$ and $\Delta^-$ need three light
quarks of one flavor, and for this fraction the factor $L_2L_1$ is replaced by
$(L_3+L_2L_1)/2$. About 30\% of the charged pions come from $\rho^0$, $\omega$ and $\eta$,
which contain a light quark and antiquark of the same flavor and, like the $\phi$, have
the factor $L_2+w(1-L_2)$ instead of $L_1^2$. With these fractions the ratio $p/\pi$
changes from 0.894 to 0.868 at $N=4$, from 0.940 to 0.924 at $N=10$ and from 0.973 to
0.966 at $N=36$. Resonance decays change the multiplicity dependence of the
proton-to-pion ratio by less than 3\%.

\subsection{Deuterons}\label{sec:nuclei}

\begin{figure*}[t]
\centering
\includegraphics[width=\widefig]{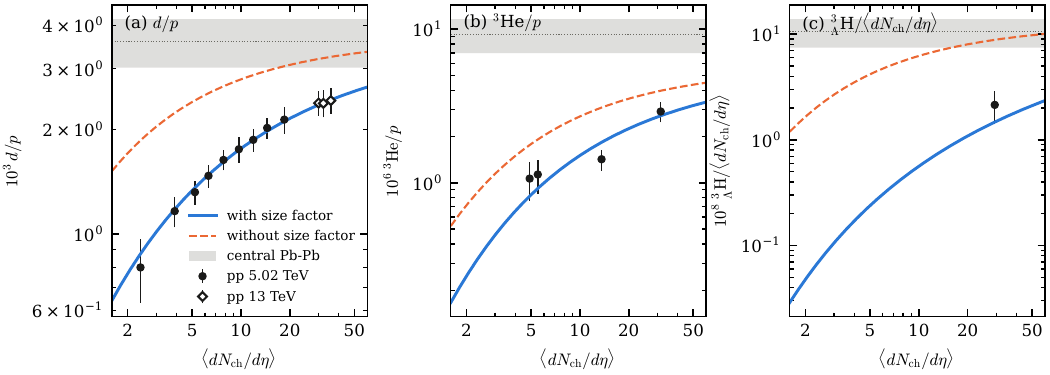}
\caption{(a) Deuteron-to-proton ratio, (b) $\He$-to-proton ratio, (c) hypertriton
yield per unit of charged-particle density, versus charged-particle multiplicity. Data:
ALICE~\cite{ALICE:2021mfm,ALICE:2021ovi,ALICE:2021sdc}. Solid lines: model with the size
factor of Eq.~(\ref{eq:G}); dashed lines: same $r$ and normalization, without the size
factor. Shaded bands:
central Pb--Pb values with their uncertainties~\cite{ALICE:2015wav,ALICE:2015oer}.}
\label{fig:nuclei}
\end{figure*}

Relative to the proton, the deuteron suppression factor is $L_3^2/(L_2L_1)$, multiplied by the size factor $G_2$. The two
effects are comparable: at $N=4.3$ quark counting gives 0.63 and the size factor 0.54,
at $N=36$ they give 0.89 and 0.75. The total suppression goes from 0.24 at $N=2.4$ to
0.67 at $N=36$. The twelve $d/p$ points in pp collisions at 5.02 and 13 TeV are
described with $\chi^2=0.5$ [Fig.~\ref{fig:nuclei}(a)], with normalization 1.02,
i.e.\ the data extrapolate to the central Pb--Pb value. 

The rise of $d/p$ with
multiplicity, by a factor four, is thus explained with no parameter specific to the
deuteron.

\subsection{\texorpdfstring{$\He$}{3He} and hypertriton}

For $\He$, with content $(5,4,0)$, quark counting gives 0.31 and the size factor 0.47 at
$N=5$, and 0.70 and 0.69 at $N=31.5$; the total suppression relative to protons is 0.15
and 0.48. The four points are described with $\chi^2=4.8$
[Fig.~\ref{fig:nuclei}(b)]; the point at $N=13.6$ lies $1.8\sigma$ below the curve. The
normalization is 0.61, $2\sigma$ below the central Pb--Pb value.

The hypertriton has content $(4,4,1)$. At $N=29.4$ quark counting gives 0.70, but the size factor is 0.17, because the
bound state is much larger than the cluster. With the large-cluster limit the model
gives a yield per unit of charged-particle density of $1.3\times10^{-8}$, against the
measured $(2.1\pm0.7)\times10^{-8}$ [Fig.~\ref{fig:nuclei}(c)].

\subsection{Predictions for nuclei}

The model makes two predictions for nuclei that have not yet been measured as functions
of multiplicity.

 The first concerns the double ratio $(\HL/\He)/(\Lambda/p)$, which
compares the hypertriton with a nucleus of the same mass number. Quark counting alone
would increase it in small clusters, since the hypertriton needs one light quark fewer
than $\He$; the size factor, however, acts in the opposite direction and dominates,
because the hypertriton is much larger than $\He$. The double ratio is therefore reduced
with respect to its large-cluster value, by a factor 0.14 at $N=4$ and 0.26 at $N=30$.

The second concerns ${}^4$He, which needs six light quarks of each flavor and has a size
parameter $r_A\simeq2.0$ fm: relative to protons it is suppressed by a factor 0.14 at $N=10$
and 0.34 at $N=30$.

\section{Cluster size and the light-to-strange ratio}\label{sec:cluster}

Table~\ref{tab:cluster} gives, for several multiplicities, the parameters of the cluster
and the main suppression factors. 

The cluster radius, obtained from $x_l$ with a light
constituent quark mass of 300 MeV, grows from 2.0 fm at $N=2.3$ to 3.7 fm at $N=45$, and the
hadronization time obtained from Eq.~(\ref{eq:d}) with $\tau_q=1$ fm from 3.2 to 4.5 fm.
This determination relies on the nuclei: the size factor of Eq.~(\ref{eq:G}) compares
the radius of the cluster with the size of the bound state, which is fixed by its wave
function, whereas the strange hadrons depend only on $x_s$ in Eq.~(\ref{eq:x}).

\begin{table*}[htb]
\caption{Cluster parameters and suppression factors versus multiplicity, with
Eqs.~(\ref{eq:fitS}) and (\ref{eq:fitr}). $R$ is the cluster radius and $\tau_h$ the
hadronization time ($\tau_q=1$ fm); $G_d$ is the size factor of the deuteron; the last
three columns are the suppression of $p/\pi$, $d/p$ and $\He/p$ relative to the
large-cluster limit.}
\label{tab:cluster}
\begin{tabular}{rrrcccccccc}
\toprule
$N$ & $x_s$ & $x_l$ & $R$ (fm) & $\tau_h$ (fm) & $L_1$ & $L_2$ & $G_d$ & $p/\pi$ & $d/p$ &
$\He/p$ \\
\midrule
2.3 & 2.6 & 9.8 & 1.96 & 3.19 & 0.95 & 0.81 & 0.47 & 0.85 & 0.23 & 0.055 \\
4.3 & 3.9 & 14.5 & 2.24 & 3.42 & 0.96 & 0.87 & 0.54 & 0.90 & 0.34 & 0.13 \\
9.8 & 6.5 & 24.4 & 2.66 & 3.75 & 0.98 & 0.92 & 0.63 & 0.94 & 0.48 & 0.26 \\
16.1 & 8.9 & 33.3 & 2.95 & 3.96 & 0.98 & 0.94 & 0.68 & 0.96 & 0.56 & 0.36 \\
23.2 & 11.2 & 41.9 & 3.19 & 4.13 & 0.99 & 0.95 & 0.71 & 0.96 & 0.61 & 0.42 \\
30.5 & 13.3 & 49.8 & 3.37 & 4.27 & 0.99 & 0.96 & 0.74 & 0.97 & 0.65 & 0.47 \\
45.0 & 16.9 & 63.7 & 3.66 & 4.46 & 0.99 & 0.97 & 0.77 & 0.98 & 0.69 & 0.54 \\
\bottomrule
\end{tabular}
\end{table*}

The corresponding volume depends on the mass of the strange quark. With $T=155$ MeV and
the constituent mass $m_s=460$ MeV, as suggested by the ratio $m_s/m_u\simeq1.5$ obtained
from the baryon magnetic moments~\cite{ParticleDataGroup:2024cfk} and by
$m_s/m_u\simeq1.5$--1.6, with $m_s-m_u\simeq173$ MeV for both mesons and baryons, from a
fit to the hadron spectrum~\cite{Karliner:2016zzc}, the values of
$x_s$ of Eq.~(\ref{eq:fitS}) give a radius from 1.5 fm at $N=2.3$ to 2.9 fm at $N=45$ and
a hadronization time from 2.8 to 3.9 fm. 

To check whether this is compatible with the
nuclei, we have repeated the fit of Sec.~\ref{sec:fit} with the size factor computed with
this radius. The fit is as good as the main one, $\chi^2=25.6$ for 42 degrees of freedom,
with $r=5.8\pm1.3$ and the normalization of $d/p$ equal to 1.14. The nuclei fix the radius
only together with $r$, and a radius about 20\% smaller is compensated by a weaker
suppression from the light quarks. 
The $\chi^2$ is nearly independent of the strange
quark mass between 300 and 500 MeV. 

The size of the cluster obtained
from strangeness with the constituent strange quark mass is therefore compatible with
the one required by the nuclei.

The value $r=3.76$ corresponds to $1/r=0.27$: a cluster contains about 0.27 strange
quarks per up (or down) quark; with the radius obtained from strangeness one finds
$1/r=0.17\pm0.04$. Both values are close to the strangeness suppression factor of string
fragmentation, $\lambda_s$, which fixes the relative probability of creating an $s\bar s$
pair rather than a $u\bar u$ or $d\bar d$ pair when a string
breaks~\cite{Andersson:1983ia}. The standard tunes of PYTHIA use
$\lambda_s\simeq0.2$~\cite{Skands:2014pea,Bierlich:2022pfr}, and larger effective
values are obtained in high-multiplicity events when the interaction of overlapping
strings is taken into account~\cite{Bierlich:2014xba}.

\section{Discussion and conclusions}\label{sec:disc}

The proposed description has a simple structure. Indeed,  one function of multiplicity, the mean number
of strange quarks per cluster, determined by the strange hadrons, and one constant, the
ratio of light to strange quark abundances, determined by the nuclei, fix the
multiplicity dependence of all the hadrons and nuclei considered.

Protons are nearly unsuppressed because the light-quark parameter $x_l$ is
large; nuclei are strongly suppressed because they need many light quarks of the same
flavor in the same cluster, and because they are as large as the cluster.

For the deuteron, the size factor of Eq.~(\ref{eq:G}) is the one of coalescence
models~\cite{Scheibl:1998tk,Bellini:2018epz}. The difference is that here the source
radius is not a free function of multiplicity but the radius of the causal cluster.

Two points require further study. (i) The normalization of $\He$ is two standard
deviations below the central Pb--Pb value, and the protons per charged particle at 13 TeV
lie below those at 5.02 TeV; both could indicate a dependence on the collision energy that
is not included here. (ii) Resonance decays are included only roughly; their effect
on protons is small, and a comprehensive assessment of feed-down would yield more precise estimates.

In conclusion, exact flavor conservation in a causal cluster, imposed on the quarks before
hadronization, extends from strangeness to the light flavors with one new constant, the
ratio of light to strange quark abundances. Together with the finite size of the nuclei,
it describes the multiplicity dependence of protons, deuterons, $\He$ and hypertriton in
pp and p--Pb collisions at the LHC. 

The model predicts the
multiplicity dependence of ${}^4$He and of the double ratio $(\HL/\He)/(\Lambda/p)$, and,
from the strangeness sector, a decrease of the $\Lambda\bar\Lambda$ spin correlation with
multiplicity proportional to the suppression of the $\phi$.

\end{document}